\documentclass[aps,pre,preprint,showpacs,showkeys,groupedaddress]{revtex4}
\newcommand{\W}{\columnwidth}
\usepackage{graphicx}
\usepackage{amsmath}
\usepackage{color}
\begin{document}
\title{Lattice-Boltzmann Method for Non-Newtonian Fluid Flows}
\author{Susana Gabbanelli}
\affiliation{Departamento de Matem\'atica y Grupo de Medios Porosos,
Facultad de Ingenier\'{\i}a, Universidad de Buenos Aires}
\author{German Drazer}
\email{drazer@mailaps.org}
\author{Joel Koplik}
\email{koplik@sci.ccny.cuny.edu}
\affiliation{Benjamin Levich Institute and Department of Physics,
City College of the City University of New York, New York, NY 10031}
\date{\today}
\begin{abstract}
We study an {\it ad hoc} extension of the Lattice-Boltzmann method
that allows the simulation of non-Newtonian fluids described by 
generalized Newtonian models. We extensively test the accuracy of the
method for the case of shear-thinning and shear-thickening truncated
power-law fluids in the parallel plate geometry, and show that the
relative error compared to analytical solutions decays approximately
linear with the lattice resolution. Finally, we also tested the method
in the reentrant-flow geometry, in which the shear-rate is no-longer a 
scalar and the presence of two singular points requires high accuracy 
in order to obtain satisfactory resolution in the local stress near 
these points. In this geometry, we also found excellent agreement with 
the solutions obtained by standard finite-element methods, 
and the agreement improves with higher lattice resolution. 
\end{abstract}
\pacs{47.11+j, 47.50.+d, 47.10.+g, 02.70.Rr}
\keywords{Lattice-Boltzmann, non-Newtonian, power-law}
\maketitle
\section{Introduction}
\label{intro}

Since its origin, more than 15 years ago, the Lattice Boltzmann Method (LBM) has proved to be
a powerful numerical technique for the simulation of single and multi-phase fluid flows in complex 
geometries. In fact, the LBM has been successfully applied to different problems in fluid dynamics 
and the interest on the method has grown rapidly in recent years. The LBM
is particularly suited for complex geometries and interfacial dynamics, and its initial 
applications included transport in porous media, multiphase and multicomponent fluid flows \cite{ChenD98}.
It was then adapted by Ladd and others to simulate particle-fluid suspensions \cite{LaddV01}.
It has also been applied to high Reynolds number incompressible flows and turbulence, and the
implementation of thermal and compressible schemes is being actively pursued \cite{YuMLW03}. 
One advantage of the LBM is that data communications between nodes is always local, which makes the
method extremely efficient for large-scale, massively-parallel computations (see Ref. \cite{NourgalievDTJ03} 
for an interesting discussion on the LBM capabilities compared to the existing continuum-based computational 
fluid dynamics methods). Another property of the LBM that has lately attracted considerable attention is the 
microscopic origin of its mesoscopic kinetic equations, which could therefore
readily incorporate molecular level interactions. This makes the LBM very compelling for microscale fluid 
dynamics in microfluidic devices \cite{Raabe04} which typically present non-continuum 
and surface-dominated effects (e.g. high Knudsen number conditions, electrokinetic and wetting phenomena).
This microscopic based
approach also makes the LBM a good candidate for hybrid, multi-scale simulations of fluid flows.

In contrast, the extension of the LBM to non-Newtonian fluids has received limited attention so far, 
in spite of the fact that a reliable extension to the LBM to simulate non-Newtonian flows would be 
very valuable; for instance, in studies of transport in geological porous media, an area in which
the LBM has been extensively applied \cite{DrazerK00,DrazerK01,DrazerK02}
due to its simple implementation in complex geometries, 
In addition to geological systems,
the flow of non-Newtonian fluids is commonly found in many areas of science and technology.

In this work, we study an {\it ad hoc} modification of the LBM, first introduced
by Aharonov and Rothman \cite{AharonovR93}, in which the local value of the viscosity depends on the 
strain-rate tensor. We show that this modification to the LBM accurately describes the flow of truncated 
power-law fluids, both shear-thinning and shear-thickening, not only in unidirectional flows (parallel 
plates geometry) but also in two-dimensional flows with simultaneous shear components in more than one 
direction (reentrant corner geometry).

\section{Lattice-Boltzmann Method}
\label{lbm}
The LBM can be viewed as an implementation of the
Boltzmann equation on a discrete lattice and for a 
discrete set of velocity distribution functions \cite{RothmanZ},
\begin{equation}
f_i(\mathbf{x+e}_i\Delta x,t+\Delta t)=f_i(\mathbf x,t) + \Omega_i(f(\mathbf x,t)),
\end{equation}
where $f_i$ is the particle velocity distribution function along the $i$th direction,
$\Omega_i(f(\mathbf x,t))$ is the {\it collision operator} which takes into account
the rate of change in the distribution function due to collisions, $\Delta x$ and
$\Delta t$ are the space and time steps discretization, respectively \cite{ChenD98}.
Then, the density $\rho$ and momentum density $\rho \mathbf u$ are given by the first
two moments of the distribution functions,
\begin{equation}
\rho = \sum_i f_i, \qquad
\rho \mathbf u = \sum_i f_i \mathbf e_i,
\end{equation}
where we assumed that the discretization is consistent with the Boltzmann equation,
$\mathbf{x+e}_i$ corresponding to the nearest neighbors of the point $\mathbf x$. 
Note that in the previous equation, and in the reminder of the article, all quantities 
are rendered dimensionless using $\Delta x$ and $\Delta t$ as the characteristic space 
and time scales, respectively. Also note that, as we are concerned with incompressible 
flows, we do not need to introduce a dimension of mass.

\subsection{BGK Approximation}
Assuming that the system is close to equilibrium the collision operator is
typically linearized about a local equilibrium distribution function,
$f_i^{eq}$, and assuming further that the local particle distribution
relaxes to equilibrium with a single characteristic (relaxation) time $\tau$,
we arrive at the Bhatnagar, Gross and Krook (BGK) approximation of the LBM \cite{ChenD98},
\begin{equation}
f_i(\mathbf{x+e}_i\Delta x,t+\Delta t)=f_i(\mathbf x,t) + 
\frac{f_i(\mathbf x,t)-f_i^{eq}(\mathbf x,t)}{\tau},
\end{equation}
where the relaxation time $\tau$ is directly related to the kinematic
viscosity of the fluid, $\nu=(2\tau-1)/6$. 

\subsection{Non-Newtonian flows}
\label{nnf}

The {\it ad hoc} extension of the LBM proposed by Aharonov and Rothman \cite{AharonovR93} to simulate
non-Newtonian fluids consists of determining the value of the relaxation time $\tau$ locally, 
in such a way that the desired local value of the viscosity is recovered.
The viscosity is related to the local rate-of-strain through the constitutive equation for the 
stress tensor. A commonly used model of non-Newtonian fluids is the {\it generalized Newtonian} 
model, in which the relation between the stress tensor, $\sigma_{ij}$, and the rate-of-strain 
tensor, $D_{ij}$, is similar to that for Newtonian fluids, $\sigma_{ij}=2\mu D_{ij}$, but with $\mu$ a 
function of the invariants of the local rate-of-strain tensor, $\mu=\mu(D_{ij})$.
In particular, we are interested in a widely used model: the power-law expression \cite{BirdSL},
$\mu=m \dot\gamma^{n-1}$, where $n>0$ is a constant characterizing the fluid. The case
$n<1$ correspond to shear-thinning (pseudoplastic) fluids, whereas $n>1$ correspond to
shear-thickening (dilatant) fluids, and $n=1$ recovers the Newtonian behavior. The magnitude of the
local shear-rate $\dot\gamma$ is related to the second invariant of the rate-of-strain tensor,
$\dot\gamma = \sqrt{D_{ij} D_{ij}}$, where the components of the rate-of-strain tensor, $D_{ij}$,
are computed locally from the velocity field. In particular, after obtaining the instantaneous velocity
field from the LBM we then compute $D_{ij}$ from a first-order finite-difference approximation to the
local derivatives of the velocity.

However, there is an obvious obstacle to a direct implementation of the power-law fluid in the LBM, 
in that the effective viscosity diverges for zero shear rates ($\dot\gamma=0$) in a
shear-thinning fluid ( $n<1$). Analogously, the viscosity becomes zero for a shear-thickening
fluid at zero shear rates. In previous studies it is not clear how this problem was avoided.

Clearly, both limits are unphysical and, in fact, it is known that many non-Newtonian fluids
exhibit a power-law behavior only in some range of shear-rates, and a constant viscosity is
observed outside that range \cite{BirdSL}. Here, we used the simplest model of such fluids:
the {\it truncated power-law} model,
\begin{equation}
\label{visc}
\nu(\dot\gamma)=\mu(\dot\gamma)/\rho  = \left\{ 
\begin{array}{ccl} 
m \dot\gamma_0^{(n-1)} &\hspace{0.5cm}& \dot\gamma< \dot\gamma_0 \\ 
m \dot\gamma^{(n-1)} && \dot\gamma_0 < \dot\gamma < \dot\gamma_{\infty}  \\ 
m \dot\gamma_{\infty}^{(n-1)} && \dot\gamma>\dot\gamma_{\infty}
\end{array} \right.
\end{equation}

Using the truncated power-law model has an additional advantage in the LBM. It is well known that
the LBM can accurately simulate viscous flows only in a limited range of kinematic viscosities.
The method becomes unstable for relaxation times close to $\tau \gtrsim 1/2$ \cite{NiuSCW04} 
(small kinematic viscosities, $\nu \lesssim 0.001$) and its accuracy is very poor 
for $\tau \gtrsim 1$ \cite{BehrendHW94} (relatively large kinematic viscosities, $\nu \gtrsim 1/6$).
Therefore, we set the lower and upper saturation values
of the kinematic viscosity in Eq.~(\ref{visc}) to $\nu_{min}= 0.001$ and 
$\nu_{max}= 0.1$. 
It is clear that the maximum value of the viscosity corresponds to the value at zero shear
rate for shear-thinning fluids ($n<1$), whereas the opposite is true for shear-thickening
materials ($n>1$).
Note that setting the value of the maximum model viscosity to $\nu_{max}=0.1$ for a given
maximum fluid viscosity $\nu^{\star}_{max}$ and a spatial resolution $\Delta x$ simply 
corresponds to choosing a particular value of the time 
step in order to satisfy, $\nu^{\star}_{max}=\nu_{max} (\Delta x^2/\Delta t)$ \cite{NourgalievDTJ03}.
Since the kinematic viscosity scales with $\Delta x^2/\Delta t$, to keep the dimensionless viscosity
constant we shall rescale $\Delta t$ according to the previous relationship 
when we increase the number of lattice nodes $N$, that is, since $\Delta x \propto 1/N$ then $\Delta t \propto 1/N^2$.
In what follow we use the three-dimensional (face-centered-hypercubic) FCHC-projection model of the LBM
with 19 velocities (D3Q19 following the notation in Ref.~\cite{QianDL92}).

\section{Flow between parallel plates}
\label{heleshaw}

We first test the proposed LBM for non-Newtonian flows in a simple unidirectional flow,
the flow between two parallel plates separated a distance $b$ in the $z$-direction 
(Hele-Shaw cell) in the presence of a pressure gradient in the $x$-direction. 
We use periodic boundary conditions in both $x$ and $y$ directions. The resulting flow field is
unidirectional, with $v_x(z)$ the only non-zero velocity component, the rate-of-strain is
a scalar function of the local velocity, $\dot\gamma=|dv_x/dz|$, and the
Navier-Stokes equations are greatly simplified \cite{Leal}.

\begin{figure}[htb]
\includegraphics*[width=\W]{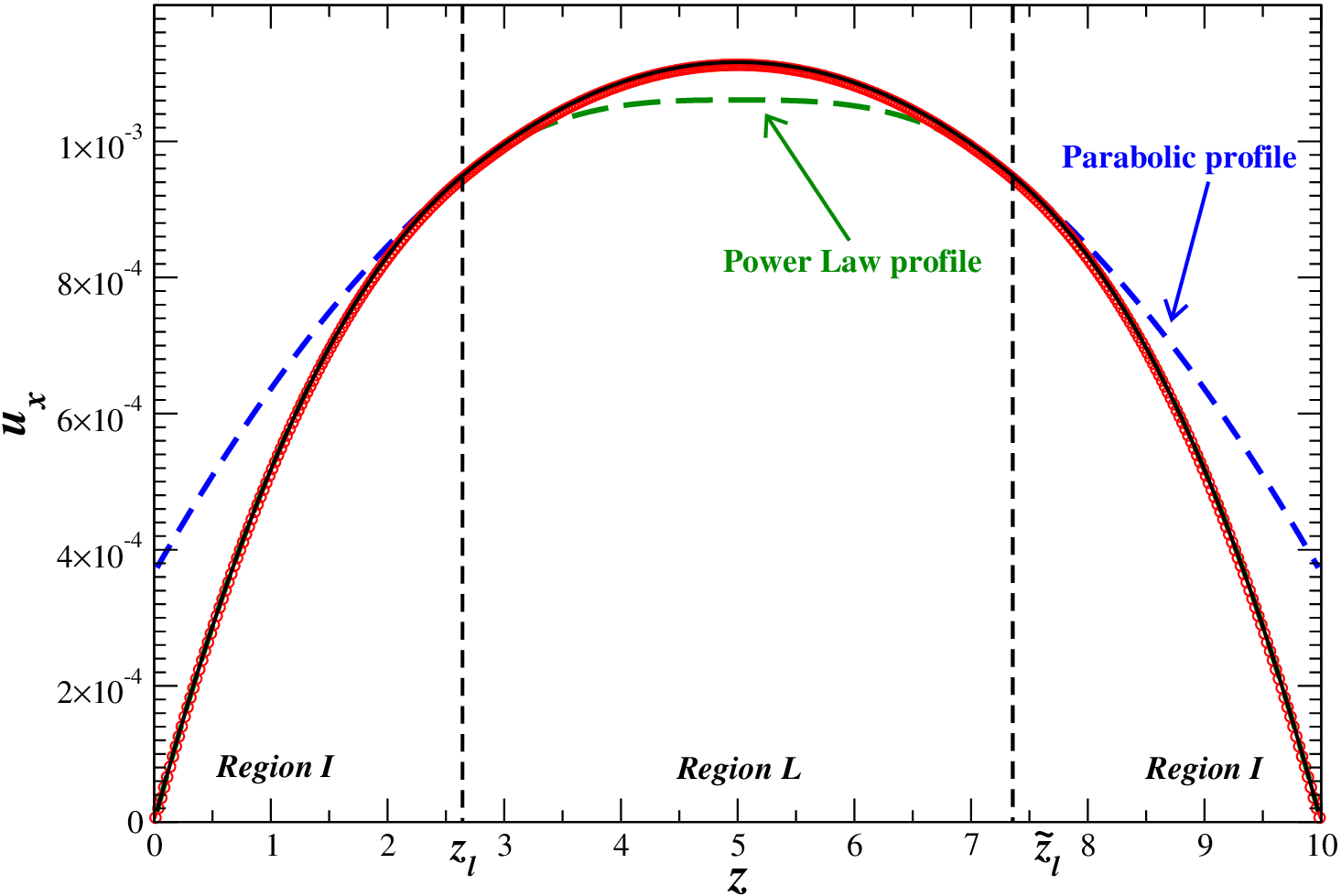}
\caption{\label{n050} Comparison between a Lattice-Boltzmann simulation and the analytical solution for the
flow between two parallel plates separated a distance $b=10$. The power-law exponent of the fluid is $n=0.50$
(shear-thinning). The pressure gradient is $\nabla P=6\times10^{-6}$, $\rho=1$, $\nu_0=0.1$, 
$\nu_{\infty}=0.001$,  $m=10^{-3}$, and $N=400$. The circles correspond to the Lattice-Boltzmann simulations.
The solid line corresponds to the analytical solution given by Eq.~(\ref{sol}).  Also shown, in dashed lines, 
are the continuation of the Newtonian and Power-Law solutions outside their regions of applicability. 
The vertical, dashed lines correspond to the transition points, $z_l$ and $\tilde z_l=b-z_l$, 
between the low shear-rate region $L$, and 
the region of intermediate shear rates, $I$. }
\end{figure}

In order to compute the exact solution to the Navier-Stokes equations for a pressure driven flow of a
truncated power-law fluid in the Hele-Shaw geometry we split the system into (in principle)
three different regions. We shall describe the regions between $z=0$ and $z=b/2$ and the analogous 
regions for $z>b/2$ follow by symmetry. The first region we consider is the high-shear rate region close 
to the walls, {\it Region H} for $z<z_h$, in which the shear rate exceeds $\dot\gamma_{\infty}$, and the 
fluid is Newtonian with effective kinematic viscosity $\nu_{\infty}=m \dot\gamma_{\infty}^{(n-1)}$; 
the second one is the intermediate region in which the fluid behaves as a power law 
according to Eq.~(\ref{visc}), {\it Region I} for $z_h<z<z_l$; and the last one is the low-shear rate region 
close to the center of the channel, {\it Region L} for $z_l<z<b/2$, 
in which the shear rate is lower than $\dot\gamma_0$ and the fluid is again 
Newtonian, but with kinematic viscosity $\nu_0=m\dot\gamma_0^{(n-1)}$.
Matching then the solution obtained in each region with the conditions of continuity in the velocity and
the stress, we obtain the general solution to the problem, in terms of the pressure gradient $G \rho =-\nabla P$,
\begin{equation}
\label{sol}
v_x(z) = \left\{
\begin{array}{lcl} 
\left( \frac{G}{2 \nu_{\infty}} \right) z (b-z)&\hspace{0.5cm} & 0 \leq z \leq z_h \\
\frac{n}{n+1} \left( \frac{G}{m} \right)^{\frac{1}{n}} 
\left[\left( \frac{b}{2} \right)^{\frac{n+1}{n}}-\left(\frac{b}{2}-z\right)^{\frac{n+1}{n}}\right]+\alpha_1
&\hspace{0.5cm}& z_h \leq z \leq z_l \\
\left(\frac{G}{2 \nu_0}\right) z (b-z)+\alpha_2 &\hspace{0.5cm}& z_l \leq z \leq b/2
\end{array} \right.,
\end{equation}
with the constants $\alpha_1$ and $\alpha_2$ given by,
\begin{eqnarray}
\alpha_1 &=& \left( \frac{G}{2 \nu_{\infty}}\right) z_h (b-z_h)- 
\frac{n}{n+1} \left( \frac{G}{m} \right)^{\frac{1}{n}}
\left[\left(\frac{b}{2}\right)^{\frac{n+1}{n}}-\left(\frac{b}{2}-z_h\right)^{\frac{n+1}{n}}\right], \\ \nonumber
\alpha_2&=&\frac{n}{n+1} \left( \frac{G}{m} \right)^{\frac{1}{n}}
\left[ \left( \frac{b}{2}\right)^{\frac{n+1}{n}}-\left(\frac{b}{2}-z_l\right)^ {\frac{n+1}{n}} \right]
-\left( \frac{G}{2 \nu_0} \right) z_l (b-z_l)+\alpha_1,
\end{eqnarray}
and the transition points, $z_h$ and $z_l$,
\begin{eqnarray}
\label{z}
z_h&=&\frac{b}{2}-\left(\frac{\nu_{\infty}}{m^{1/n}} \right)^{\frac{n}{n-1}} \frac{1}{G}
=\frac{b}{2}-\frac{m\dot\gamma_{\infty}^n}{G}, \\ \nonumber
z_l&=&\frac{b}{2}-\left(\frac{\nu_{0}}{m^{1/n}} \right)^{\frac{n}{n-1}} \frac{1}{G}
=\frac{b}{2}-\frac{m\dot\gamma_0^n}{G}.
\end{eqnarray}
Clearly, the number of regions that coexist will depend on the magnitude of the imposed pressure 
gradient $G$. For very small pressure gradients, $G\ll1$, both $z_h$ and $z_l$ become negative 
(see the previous equation), which means that shear rates are smaller than $\dot\gamma_0$ across 
the entire gap and only {\it Region L} exists. As $G$ increases, 
there is a range of pressure gradients, $m\dot\gamma_0^n < (b/2) G < m \dot\gamma_{\infty}^n$, 
for which $z_l>0$ but $z_h<0$, and therefore regions {\it L} and {\it I} coexist. Finally, 
for $G>(2/b)m\dot\gamma_{\infty}^n$ we obtain $z_h>0$, and all three
regions are present in the flow. Note that for large values of $G$ both transition
points converge to the center of the cell, $z_l, z_h \to b/2$.
Thus, Eq.~(\ref{z}) allows us to choose the appropriate value of $G$ in order to investigate the different 
regimes.

\begin{figure}[htb]
\includegraphics*[width=\W]{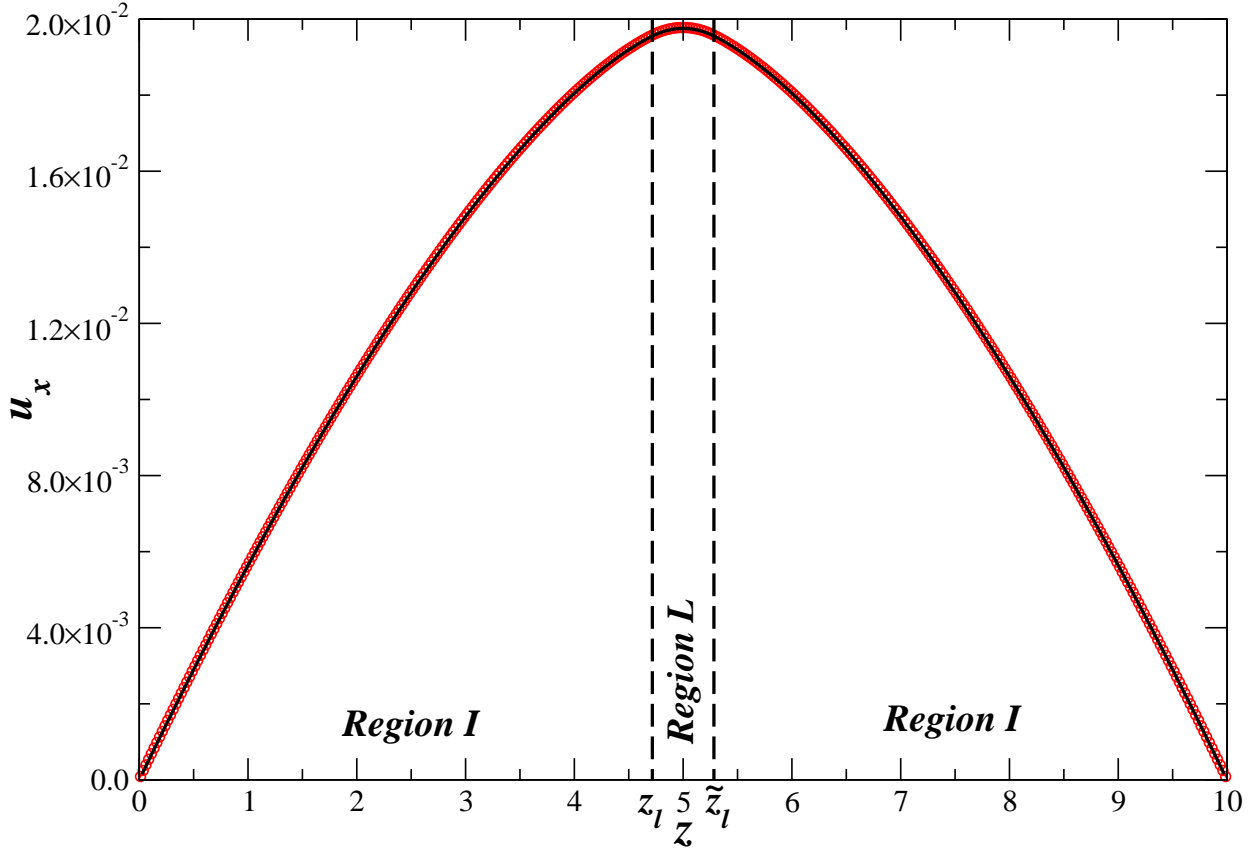}
\caption{\label{n200} Comparison between a Lattice-Boltzmann simulation and the analytical solution for the
flow between two parallel plates separated a distance $b=10$. The power-law exponent of the fluid is $n=2.00$
(shear-thickening). The pressure gradient is $\nabla P=5\times10^{-6}$, $\rho=1$, $\nu_0=0.001$, 
$\nu_{\infty}=0.1$, $m=10^{-3}$, and $N=400$. The circles correspond to the Lattice-Boltzmann simulations and
the solid line corresponds to the analytical solution given by Eq.~(\ref{sol}). 
The vertical dashed lines correspond to the transition points, $z_l$ and $\tilde z_l=b-z_l$, 
between the low shear-rate region $L$ and 
the intermediate shear rates region $I$. }
\end{figure}

We performed a large number of simulations for different values of the power-law exponent $n$. Specifically,
we consider two shear-thinning fluids, $n=0.50$ and $n=0.75$, and two shear-thickening fluids,
$n=1.25$ and $n=2.00$. In all cases we performed simulations for two different magnitudes of the
external forcing: one for which the region of low shear rates {\it L} is important, that is relatively small
pressure gradients for which $z_l \sim b/4$; and a second one in which the fluid behaves as a power-law fluid 
almost in the entire gap, that is  $z_l \sim b/2$. In both cases the shear-rate does not exceeds $\dot\gamma_{\infty}$.
In Fig.~\ref{n050} we present a comparison between the Lattice-Boltzmann results and the analytical solution given
in Eq.~(\ref{sol}) for a shear-thinning fluid with power-law exponent $n=0.50$. The simulation corresponds
to a relatively small pressure gradient for which the region of small shear-rates is large, 
$z_l\sim b/4$. Both regions, {\it Region L} in which the fluid behaves as a Newtonian one, 
and {\it Region I} in which the effective viscosity is a power-law, are shown. 
The agreement with the analytical solution is excellent, with relative error close to $0.1\%$.  
In Fig.~\ref{n200} we present a similar comparison between the LBM and the analytical solution, but for a
shear-thickening fluid ($n=2.00$) which behaves as a power-law fluid across almost the entire channel. Again
the agreement is excellent with relative error smaller than $0.1\%$.

\begin{figure}[htb]
\includegraphics*[width=\W]{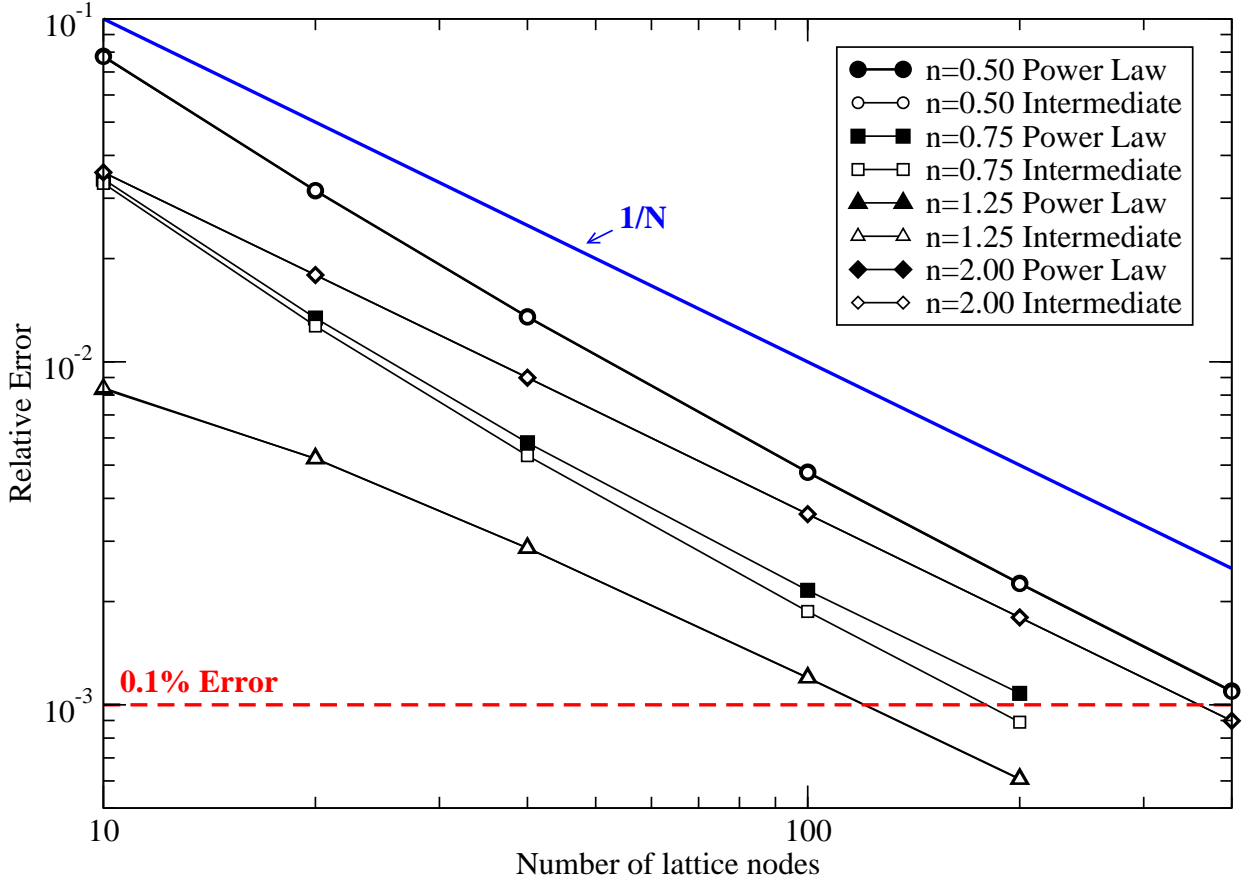}
\caption{\label{error}Relative error of the LBM compared to the analytical solution for the 
flow between parallel plates, as a function of the number of lattice points used in the simulations.
The points correspond to simulations with the LBM for four different fluids, two shear-thinning fluids 
($n=0.50$ and $n=0.75$) and two shear-thickening fluids ($n=1.25$ and $n=2.00$). For all fluids we also present
results corresponding to two different regimes: 
one at high pressure gradients, in which the low-shear-rates region, {\it region L}, 
is small (Power-Law) and the other one at intermediate pressure gradients in which both regions {\it L}  and {\it I} 
are comparable (note that with the exception of $n=0.75$ both regimes give almost exactly the same relative error and,
in fact, the corresponding points overlap almost entirely). 
In all cases, we increased the lattice resolution until the relative error was on the order of
$0.1\%$. The solid line shows the general trend of the data, $1/N$.}
\end{figure}

Finally, for each of these cases we run a series of simulations in which the number of lattice nodes, $N$, in the direction of 
the gap was increased from $10$ to $400$, and computed the relative error of the LBM results compared to the
analytical solutions, defined as $\sum_{i=1}^{N} (1-v_i^{LBM}/v_i^{Anal.})^2$. In order to obtain the accuracy of the LBM as a function
of the number of nodes, we simulated the same physical problem but changed $\Delta x$ from $1$ to $0.025$. 
In addition, since the accuracy of the LBM depends on the model viscosity, we also changed $\Delta t$ according to 
$\Delta t=\Delta x^2$, so that the model viscosity remains the same, independent of the number of nodes. 
Then, in order to compare the velocity field always at the same physical time since startup, the number of time steps was increased 
inversely proportional to $\Delta t$ (reaching $\sim 10^8$ time steps for $N=400$). In Fig.~\ref{error} we present the results
obtained for the different fluids and different pressure gradients. It is clear that, in all cases, the relative error
decreases, approximately as $1/N$, as the number of nodes is increased, and eventually becomes of the order of $0.1\%$
(an arbitrary target accuracy that we set for our simulations). The error was found to be independent of the
pressure gradient, or the size of the non-Newtonian region $I$, but strongly depends on the power-law exponent. In particular,
the relative error seems to increase as the magnitude of $(1-n)/n$ increases, with the error in the Newtonian case ($n=1$) decaying
faster than $1/N$. This is probably related to the first-order finite-difference approximation used to compute the spatial 
derivatives of the fluid velocity which determine the local viscosity through Eq.~(\ref{visc}). It would then be possible to improve
the accuracy of the method by implementing a higher order approximation of the local shear-rates.

\section{Reentrant corner flow}
\label{reentrant}

\begin{figure}[htb]
\includegraphics[width=\W]{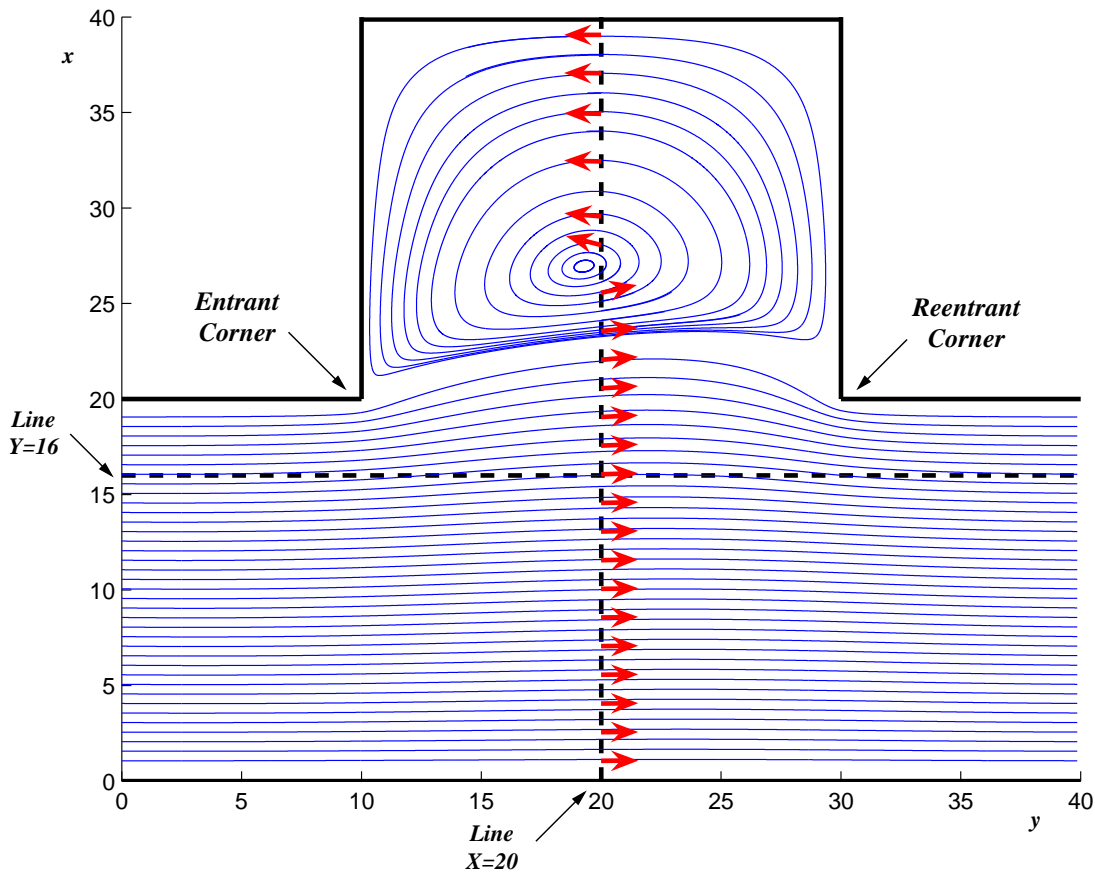}
\caption{\label{setup} Streamlines in the reentrant flow geometry. The flow direction is shown at the center of the channel.
Note the asymmetry due to inertia effects. The dashed lines show the lines in which we compare the solutions of the LBM method
with the solutions obtained by finite-element calculations.}
\end{figure}

\begin{figure}[htb]
\includegraphics[width=\W]{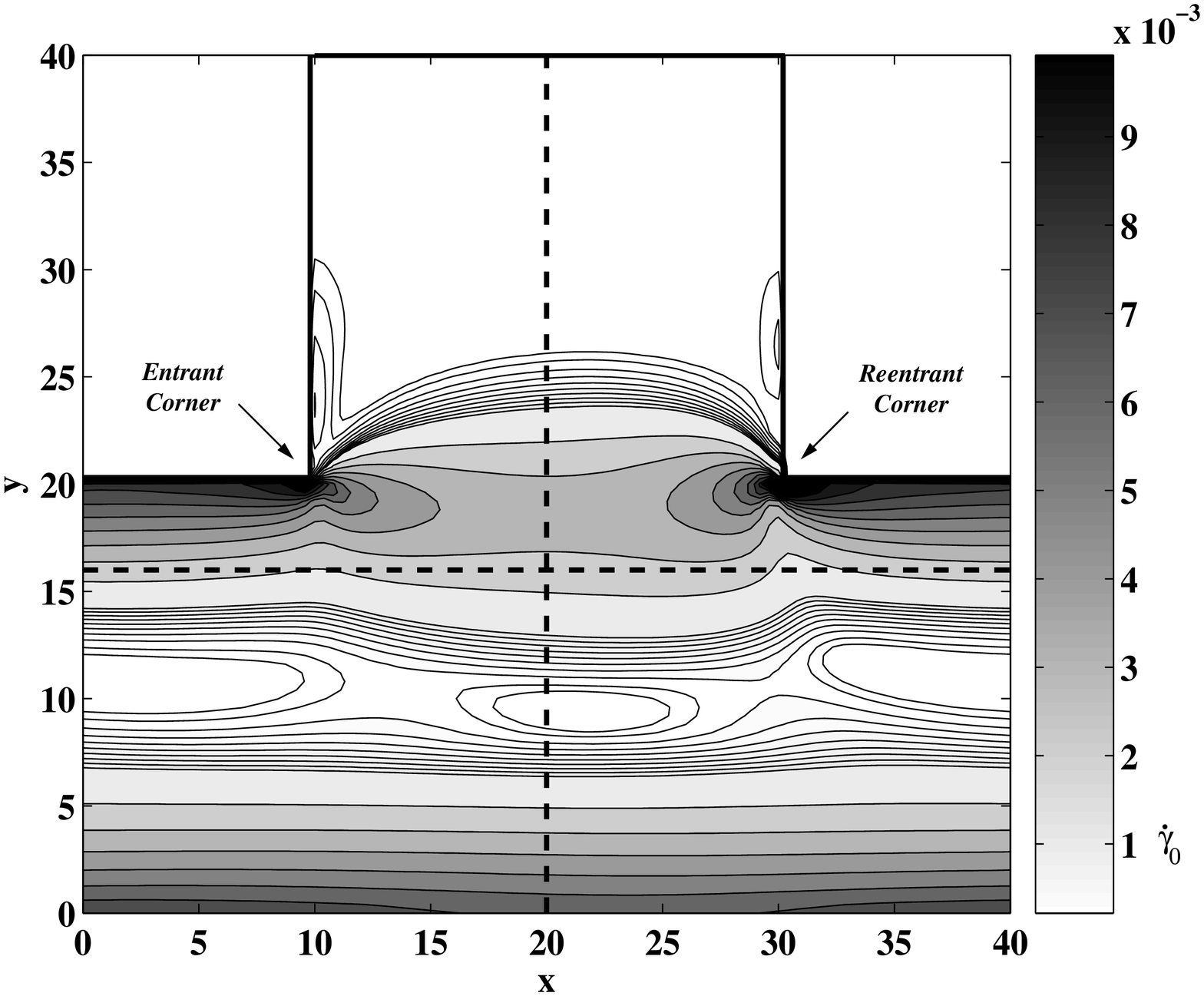}
\caption{\label{stress} Contour-plot of the local magnitude of the shear-rate, as computed with the LBM.
The lowest level of the plot corresponds to $\dot\gamma_0$, that is the fluid behaves as Newtonian
in those regions. High shear-rates, and accordingly high shear-stresses are localized at the {\it entrant} and
{\it reentrant} corners.}
\end{figure}

In the previous section we tested the LBM for non-Newtonian fluids in a Hele-Shaw geometry and
found excellent agreement with the analytical solutions as the number of nodes was increased. 
In that case, the flow is unidirectional and therefore the shear-rate is a scalar, which
is a rather simple type of flow. In contrast,  we shall now test the LBM in a more demanding geometry,
that is the {\it reentrant corner} geometry sketched in Fig.~\ref{setup}. In this case,
the shear-rate is no longer a scalar as in the Hele-Shaw geometry and, in addition, the presence of
two singular points, located at the {\it entrant} and {\it reentrant} corners (see Fig.~\ref{setup}), 
requires high accuracy in order to obtain satisfactory stress resolution near these points 
(although no analog to the Moffatt's analysis near the corner is available for non-Newtonian fluids, it is 
believed that not-integrable stress singularities develop in this case, and numerical techniques do not 
always converge \cite{KoplikB97}).
Motivated by these issues we simulated the flow in the reentrant corner geometry using the LBM
for a shear-thinning fluid with power-law exponent $n=0.50$. In Fig.~\ref{setup} we present the
streamlines corresponding to the computed velocity field, obtained for a pressure 
gradient $\nabla P=10^{-5}$, where the recirculation region inside the cavity can be observed
(note that the separation between streamlines was chosen for visualization purposes only and it is not
related to the local flow rate, since the magnitude of the fluid velocity sharply decays inside the 
cavity).  The corresponding Reynolds number is $Re=4$, computed with the maximum 
viscosity, $\nu_0$, and the measured mean velocity. In fact, the streamlines shown in Fig.~\ref{setup} are 
fore-aft asymmetric, due to inertia effects, which are absent in low-Reynolds-number flows.
We also computed the local magnitude of the shear-rate, related to the local stress field
through the constitutive relation given by Eq.~(\ref{visc}).
In Fig.~\ref{stress} we present a contour plot of the magnitude of the shear-rate in the reentrant corner 
geometry with the lowest level in the contour plot corresponding to $\dot\gamma_0$.
It can be seen that the fluid is Newtonian in small regions 
at the center of the channel and inside the recirculation region. It is also clear that, as discussed before,
both the {\it entrant} and {\it reentrant} corners are singular points where the shear-rate increases
to its highest values in the system. 

\begin{figure}[htb]
\includegraphics*[width=\W]{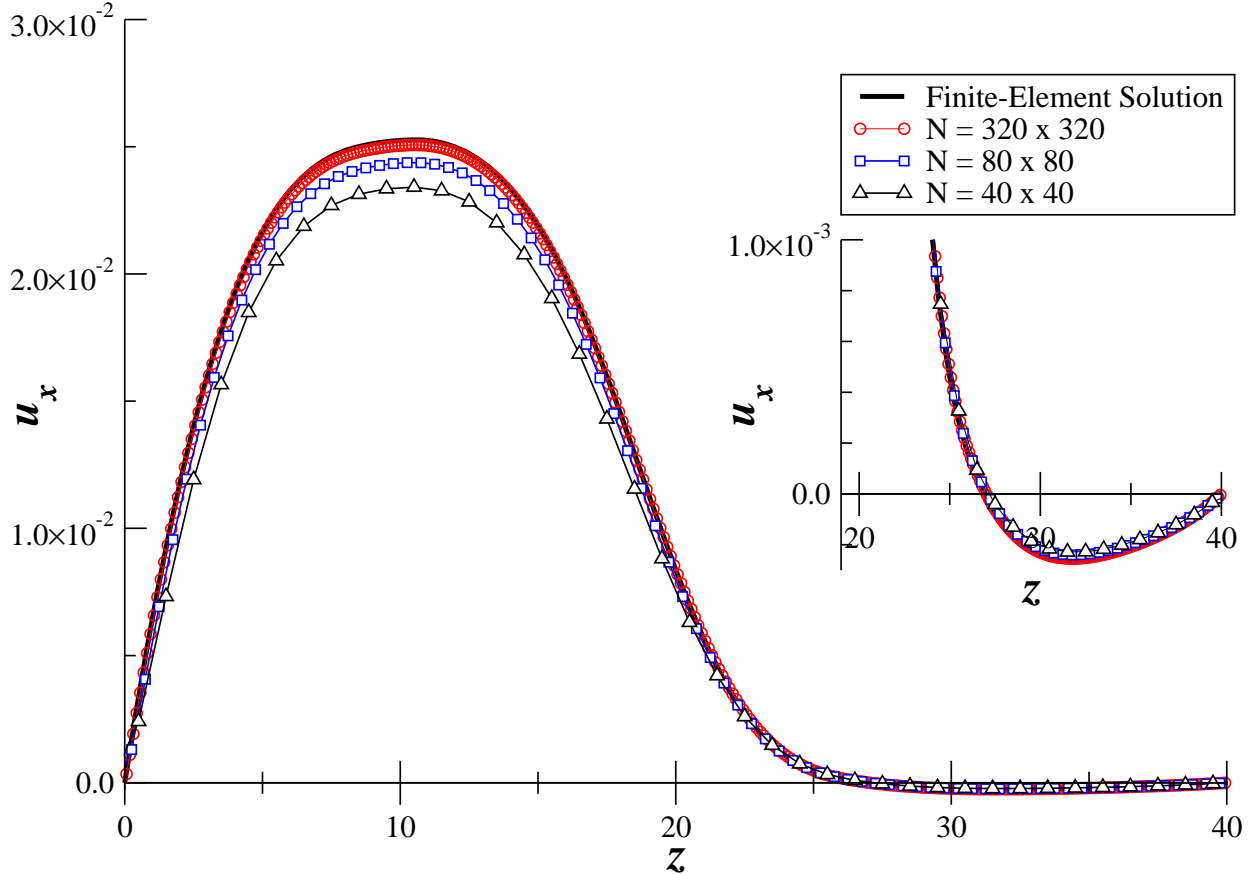}
\caption{\label{ux} Velocity component along the channel, $u_x$, plotted in a line perpendicular to the flow
(dashed line $X=20$ in Fig.~\ref{setup}). We compare the results of the finite-element calculations (solid line)
with the results of the LBM (points) for different lattice resolutions. In the inset we plot the velocity profile
inside the recirculation region.}

\end{figure}
\begin{figure}[htb]
\includegraphics*[width=\W]{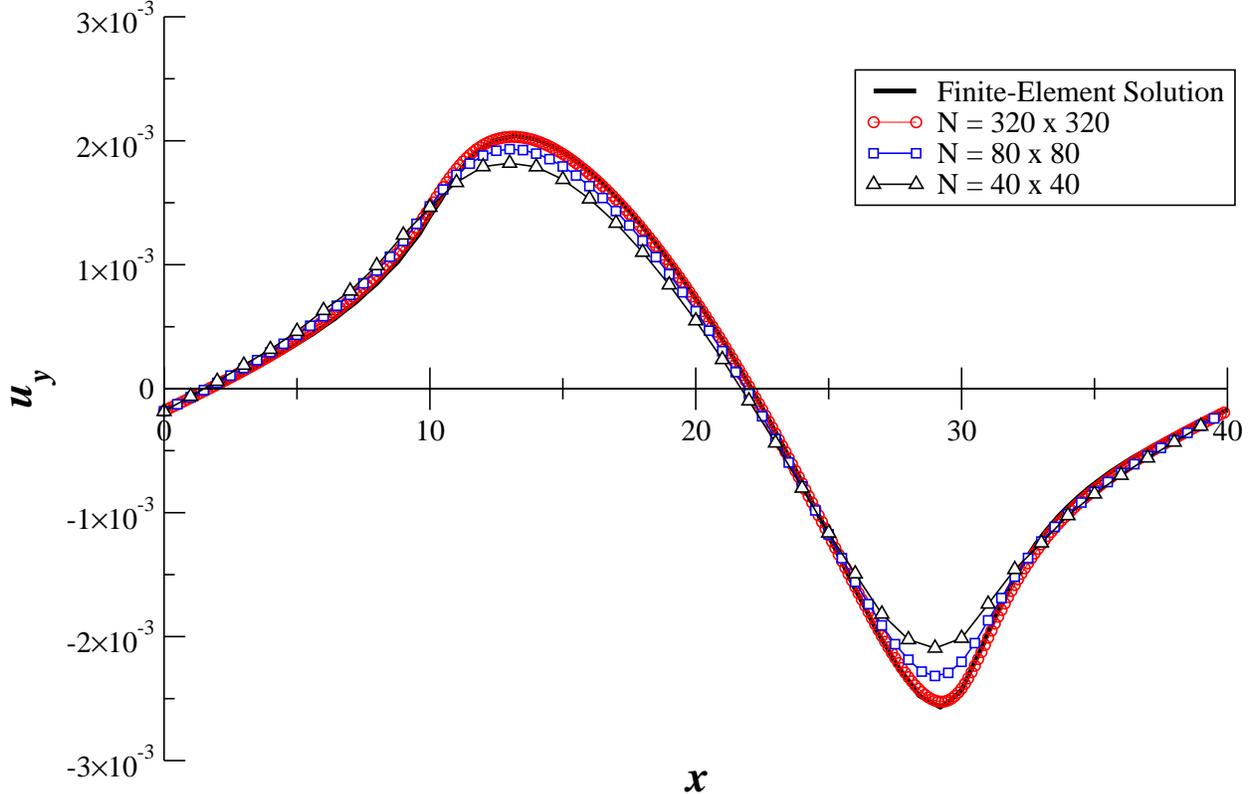}
\caption{\label{uy} Velocity component perpendicular to the flow direction, $u_y$, plotted in a line parallel
to the flow and close to the top wall of the channel (dashed line $Y=16$ in Fig.~\ref{setup}). We compare the
results of the finite-element calculations (solid line) with the results of the LBM (points) for different 
lattice resolutions.} 
\end{figure}

Finally, in order to perform a more quantitative test of the LBM, we solved the problem
numerically using the finite-element commercial software FIDAP (Fluent Inc.). 
In Figs.~\ref{ux} and \ref{uy} we compare the solutions obtained with the LBM for different
resolutions, ranging from $N=40 \times 40$ to $N=320 \times 320$, with the finite-element results 
obtained with FIDAP. The comparison is made along two lines, 
one oriented along the flow direction (dashed line at $Y=16$ in Fig.~\ref{setup}) and a second one oriented in 
the perpendicular direction (dashed line at $X=20$ in Fig.~\ref{setup}).
In Fig.~\ref{ux} we compare the velocity along the channel, $u_x$, in the line perpendicular to $x$
that is located at the center of the system ($X=20$, see Fig.~\ref{setup}).
A velocity profile similar to that in a Hele-Shaw cell is observed for $0 < y \lesssim 20$, as well
as the recirculation flow inside the cavity (see the inset). 
In Fig.~\ref{uy} we plot the velocity in the vertical direction, $u_y$, along a horizontal line 
close to the upper wall of the channel ($Y=16$, see Fig.~\ref{setup}).
It is clear that there is some fluid penetration into the cavity ({\it entrant flow})
in the first half of the channel and some {\it reentrant flow} in the second half
(note that, as mentioned before, the flow is not symmetric about $x=20$ due to inertia
effects). 
In both cases we found an excellent agreement between the two methods, and the
agreement clearly improved as the number of lattice nodes was increased in the LBM (the number of elements 
in the finite-element computations was fixed to $100 \times 100$). 

\section{Conclusions}
We have extensively tested an {\it ad hoc} modification of the Lattice-Boltzmann Method that extends
its use to Generalized Newtonian fluids, in which the non-Newtonian character of the
fluids is modelled as an effective viscosity. Specifically, we calculated the accuracy of the method
for Truncated Power-Law fluids and showed that the relative error decays (linearly) as the
resolution of the lattice (number of lattice points) is increased. The error was computed
directly from the analytical solutions of the problem. The same trend was observed for both
shear-thinning ($n<1$) and shear-thickening ($n>1$) fluids, as well as for intermediate and
high shear-rates. In all cases the relative error was of the order of $0.1\%$
for the highest resolution employed.
Finally, we also tested the method in the reentrant flow
geometry and showed that it is in excellent agreement with the solution obtained by means of finite-element
calculations. Again, the accuracy of the method was shown to increase with the number of lattice points.


\section*{Acknowledgments}

This work is part of a collaboration supported by the Office of International Science and Engineering
of the National Science Foundation under Grant No. INT-0304781.
This research was supported by the Geosciences Research Program, Office of
Basic Energy Sciences, U.S. Department of Energy, and computational facilities
were provided by the National Energy Resources Scientific
Computer Center. 

\bibliography{articles,book,mybooks,mios}

\begin{thebibliography}{16}
\expandafter\ifx\csname natexlab\endcsname\relax\def\natexlab#1{#1}\fi
\expandafter\ifx\csname bibnamefont\endcsname\relax
  \def\bibnamefont#1{#1}\fi
\expandafter\ifx\csname bibfnamefont\endcsname\relax
  \def\bibfnamefont#1{#1}\fi
\expandafter\ifx\csname citenamefont\endcsname\relax
  \def\citenamefont#1{#1}\fi
\expandafter\ifx\csname url\endcsname\relax
  \def\url#1{\texttt{#1}}\fi
\expandafter\ifx\csname urlprefix\endcsname\relax\def\urlprefix{URL }\fi
\providecommand{\bibinfo}[2]{#2}
\providecommand{\eprint}[2][]{\url{#2}}

\bibitem[{\citenamefont{Chen and Doolen}(1998)}]{ChenD98}
\bibinfo{author}{\bibfnamefont{S.}~\bibnamefont{Chen}} \bibnamefont{and}
  \bibinfo{author}{\bibfnamefont{G.~D.} \bibnamefont{Doolen}},
  \bibinfo{journal}{Ann. Rev. Fluid Mech.} \textbf{\bibinfo{volume}{30}},
  \bibinfo{pages}{329} (\bibinfo{year}{1998}).

\bibitem[{\citenamefont{Ladd and Verberg}(2001)}]{LaddV01}
\bibinfo{author}{\bibfnamefont{A.~J.~C.} \bibnamefont{Ladd}} \bibnamefont{and}
  \bibinfo{author}{\bibfnamefont{R.}~\bibnamefont{Verberg}},
  \bibinfo{journal}{J. Stat. Phys.} \textbf{\bibinfo{volume}{104}},
  \bibinfo{pages}{1191} (\bibinfo{year}{2001}).

\bibitem[{\citenamefont{Yu et~al.}(2003)\citenamefont{Yu, Mei, Luo, and
  Wei}}]{YuMLW03}
\bibinfo{author}{\bibfnamefont{D.~Z.} \bibnamefont{Yu}},
  \bibinfo{author}{\bibfnamefont{R.~W.} \bibnamefont{Mei}},
  \bibinfo{author}{\bibfnamefont{L.~S.} \bibnamefont{Luo}}, \bibnamefont{and}
  \bibinfo{author}{\bibfnamefont{S.}~\bibnamefont{Wei}},
  \bibinfo{journal}{Prog. Aerosp. Sci.} \textbf{\bibinfo{volume}{39}},
  \bibinfo{pages}{329} (\bibinfo{year}{2003}).

\bibitem[{\citenamefont{Nourgaliev et~al.}(2003)\citenamefont{Nourgaliev, Dinh,
  Theofanous, and Joseph}}]{NourgalievDTJ03}
\bibinfo{author}{\bibfnamefont{R.~R.} \bibnamefont{Nourgaliev}},
  \bibinfo{author}{\bibfnamefont{T.~N.} \bibnamefont{Dinh}},
  \bibinfo{author}{\bibfnamefont{T.~G.} \bibnamefont{Theofanous}},
  \bibnamefont{and} \bibinfo{author}{\bibfnamefont{D.}~\bibnamefont{Joseph}},
  \bibinfo{journal}{Int. J. Mult. Flow} \textbf{\bibinfo{volume}{29}},
  \bibinfo{pages}{117} (\bibinfo{year}{2003}).

\bibitem[{\citenamefont{Raabe}(2004)}]{Raabe04}
\bibinfo{author}{\bibfnamefont{D.}~\bibnamefont{Raabe}},
  \bibinfo{journal}{Modelling Simul. Mater. Sci. Eng.}
  \textbf{\bibinfo{volume}{12}}, \bibinfo{pages}{R13} (\bibinfo{year}{2004}).

\bibitem[{\citenamefont{Drazer and Koplik}(2000)}]{DrazerK00}
\bibinfo{author}{\bibfnamefont{G.}~\bibnamefont{Drazer}} \bibnamefont{and}
  \bibinfo{author}{\bibfnamefont{J.}~\bibnamefont{Koplik}},
  \bibinfo{journal}{Phys. Rev. E} \textbf{\bibinfo{volume}{62}},
  \bibinfo{pages}{8076} (\bibinfo{year}{2000}).

\bibitem[{\citenamefont{Drazer and Koplik}(2001)}]{DrazerK01}
\bibinfo{author}{\bibfnamefont{G.}~\bibnamefont{Drazer}} \bibnamefont{and}
  \bibinfo{author}{\bibfnamefont{J.}~\bibnamefont{Koplik}},
  \bibinfo{journal}{Phys. Rev. E} \textbf{\bibinfo{volume}{63}},
  \bibinfo{eid}{056104} (\bibinfo{year}{2001}).

\bibitem[{\citenamefont{Drazer and Koplik}(2002)}]{DrazerK02}
\bibinfo{author}{\bibfnamefont{G.}~\bibnamefont{Drazer}} \bibnamefont{and}
  \bibinfo{author}{\bibfnamefont{J.}~\bibnamefont{Koplik}},
  \bibinfo{journal}{Phys. Rev. E} \textbf{\bibinfo{volume}{66}},
  \bibinfo{eid}{026303} (\bibinfo{year}{2002}).

\bibitem[{\citenamefont{Aharonov and Rothman}(1993)}]{AharonovR93}
\bibinfo{author}{\bibfnamefont{E.}~\bibnamefont{Aharonov}} \bibnamefont{and}
  \bibinfo{author}{\bibfnamefont{D.~H.} \bibnamefont{Rothman}},
  \bibinfo{journal}{Geophys. Res. Lett.} \textbf{\bibinfo{volume}{20}},
  \bibinfo{pages}{679} (\bibinfo{year}{1993}).

\bibitem[{\citenamefont{Rothman and Zaleski}(1997)}]{RothmanZ}
\bibinfo{author}{\bibfnamefont{D.~H.} \bibnamefont{Rothman}} \bibnamefont{and}
  \bibinfo{author}{\bibfnamefont{S.}~\bibnamefont{Zaleski}},
  \emph{\bibinfo{title}{Lattice-Gas Cellular Automata, Simple Models of Complex
  Hydrodynamics}} (\bibinfo{publisher}{Cambridge University Press},
  \bibinfo{address}{Cambridge, UK}, \bibinfo{year}{1997}).

\bibitem[{\citenamefont{Bird et~al.}(2001)\citenamefont{Bird, Stewart, and
  Lightfoot}}]{BirdSL}
\bibinfo{author}{\bibfnamefont{R.~B.} \bibnamefont{Bird}},
  \bibinfo{author}{\bibfnamefont{W.~E.} \bibnamefont{Stewart}},
  \bibnamefont{and} \bibinfo{author}{\bibfnamefont{E.~N.}
  \bibnamefont{Lightfoot}}, \emph{\bibinfo{title}{Transport Phenomena}}
  (\bibinfo{publisher}{Wiley Text Books}, \bibinfo{address}{New York},
  \bibinfo{year}{2001}), \bibinfo{edition}{{2nd}} ed.

\bibitem[{\citenamefont{Niu et~al.}(2004)\citenamefont{Niu, Shu, Chew, and
  Wang}}]{NiuSCW04}
\bibinfo{author}{\bibfnamefont{X.~D.} \bibnamefont{Niu}},
  \bibinfo{author}{\bibfnamefont{C.}~\bibnamefont{Shu}},
  \bibinfo{author}{\bibfnamefont{Y.~T.} \bibnamefont{Chew}}, \bibnamefont{and}
  \bibinfo{author}{\bibfnamefont{T.~G.} \bibnamefont{Wang}},
  \bibinfo{journal}{J. Stat. Phys.} \textbf{\bibinfo{volume}{117}},
  \bibinfo{pages}{665} (\bibinfo{year}{2004}).

\bibitem[{\citenamefont{Behrend et~al.}(1994)\citenamefont{Behrend, Harris, and
  Warren}}]{BehrendHW94}
\bibinfo{author}{\bibfnamefont{O.}~\bibnamefont{Behrend}},
  \bibinfo{author}{\bibfnamefont{R.}~\bibnamefont{Harris}}, \bibnamefont{and}
  \bibinfo{author}{\bibfnamefont{P.~B.} \bibnamefont{Warren}},
  \bibinfo{journal}{Phys. Rev. E} \textbf{\bibinfo{volume}{50}},
  \bibinfo{pages}{4586} (\bibinfo{year}{1994}).

\bibitem[{\citenamefont{Qian et~al.}(1992)\citenamefont{Qian, D'Humieres, and
  Lallemand}}]{QianDL92}
\bibinfo{author}{\bibfnamefont{Y.~H.} \bibnamefont{Qian}},
  \bibinfo{author}{\bibfnamefont{D.}~\bibnamefont{D'Humieres}},
  \bibnamefont{and}
  \bibinfo{author}{\bibfnamefont{P.}~\bibnamefont{Lallemand}},
  \bibinfo{journal}{Europhys. Lett.} \textbf{\bibinfo{volume}{17}},
  \bibinfo{pages}{479} (\bibinfo{year}{1992}).

\bibitem[{\citenamefont{Leal}(1992)}]{Leal}
\bibinfo{author}{\bibfnamefont{L.~G.} \bibnamefont{Leal}},
  \emph{\bibinfo{title}{Laminar Flow and Convective Transport Processes}}
  (\bibinfo{publisher}{Butterworth-Heinemann}, \bibinfo{year}{1992}).

\bibitem[{\citenamefont{Koplik and Banavar}(1997)}]{KoplikB97}
\bibinfo{author}{\bibfnamefont{J.}~\bibnamefont{Koplik}} \bibnamefont{and}
  \bibinfo{author}{\bibfnamefont{J.~R.} \bibnamefont{Banavar}},
  \bibinfo{journal}{J. Rheol.} \textbf{\bibinfo{volume}{41}},
  \bibinfo{pages}{787} (\bibinfo{year}{1997}).

\end{thebibliography}
\bibliographystyle{apsrev}

\end{document}